\documentclass[twocolumn,aps,pra]{revtex4-1}

\usepackage{amssymb}
\usepackage{amsmath}    
\usepackage{verbatim}   
\usepackage{color}      
\usepackage{subfigure}  
\usepackage{epsfig}
\usepackage{xspace}

\newcommand{\dblwel}{double-well\xspace}
\newcommand{\Dblwel}{Double-well\xspace}
\newcommand{\DblWel}{Double-Well\xspace}
\newcommand{\probdist}{probability distribution\xspace}
\newcommand{\wavfun}{wave function\xspace}
\newcommand{\grndst}{ground state\xspace}
\newcommand{\frstxit}{first excited\xspace}
\newcommand{\multfun}{multiplier-function\xspace}
\newcommand{\tildeN}{\mathcal{N}\xspace}

\begin{document}
\title{\DblWel Quantum Tunneling Visualized via Wigner's Function}
\author{Dimitris Kakofengitis}
\email{D.Kakofengitis@herts.ac.uk}
\author{Ole Steuernagel}
\email{O.Steuernagel@herts.ac.uk}
\affiliation{School of Physics Astronomy and Mathematics,
University of Hertfordshire, Hatfield, AL10 9AB, UK}

\date{\today}

\begin{abstract}
  We investigate quantum tunneling in smooth symmetric and asymmetric
  \dblwel potentials. Exact solutions for the ground and \frstxit
  states are used to study the dynamics. We introduce Wigner's
  quasi-\probdist function to highlight and visualize the
  non-classical nature of spatial correlations arising in tunneling.
\end{abstract}
\maketitle
\section{Introduction}
Quantum tunneling was first discussed by Friedrich Hund in 1927 when
he considered the \grndst of a \dblwel
potential~\cite{Nimtz_08,Merzbacher_02}. Quantum tunneling is a
microscopic phenomenon where a particle can penetrate into and in
some cases pass through a potential barrier, although the barrier is
energetically higher than the kinetic energy of the particle. This
motion amounts to particles penetrating areas where they have
``negative kinetic energy'' and is not allowed by the laws of
classical dynamics~\cite{Razavy_03}. \Dblwel potentials can be used
for the study of tunneling, as the central hump separating the two
wells constitutes a tunneling barrier, for eigenstates lower in
energy than the maximum height of the barrier. They are also used for
the study of molecular structures, for example in the ammonia
molecule~\cite{Ka_03}. The concept of quantum tunneling is central
to the operation of scanning tunneling
microscopes~\cite{Bai_00,Carminati_95}.

In this paper we investigate tunneling in partially exactly solvable
\dblwel potentials~\cite{Caticha_95} through the use of the Wigner
quasi-\probdist function. We model tunneling based on the dynamics of
the lowest two \wavfun{}s, the quantum mechanical \grndst and the
\frstxit state~\cite{Caticha_95}. We also present the \probdist{}s
with respect to momentum and position, and illustrate the presence of
quantum coherences, which give rise to interference fringes with
negative values of Wigner's quasi-\probdist.

In the next section, we provide a self-consistent discussion of the
partially exactly solvable symmetric and asymmetric \dblwel
potentials, first established in~\cite{Caticha_95}. In
section~\ref{sec:WignerFunction} we introduce Wigner's
quasi-\probdist function and illustrate, for each case of the
\dblwel potential, time-evolution of the associated Wigner
quasi-\probdist, and \probdist{}s with respect to position and
momentum. Subsection~\ref{sec:QuantumCoherence} considers negative
regions of Wigner's quasi-\probdist function and allows us to
visualize non-classical spatial coherences in tunneling, finally we
conclude.

\section{Partially Exactly Solvable \DblWel Potentials}
\subsection{The Schr\"odinger Equation}\label{sec:Schrodinger}

A family of partially exactly solvable \dblwel potentials was
introduced in 1995 by Caticha~\cite{Caticha_95}; for these the
ground and \frstxit states (and in some cases states energetically
above those) can be computed analytically in simple closed forms.
Note that despite considerable efforts, only few fully solvable
smooth potentials are known~\cite{Gendenshtein_83,Cooper_95}; no
fully solvable smooth \dblwel potential has yet been found. Such
partially solvable models may well prove to be very useful as
benchmarks for computer code and numerical tests.

\begin{figure}[ht]
\centering
\subfigure{\label{fig:SymmetricSuperpotentialPotentialPhi}
\includegraphics[width=0.23\textwidth]
{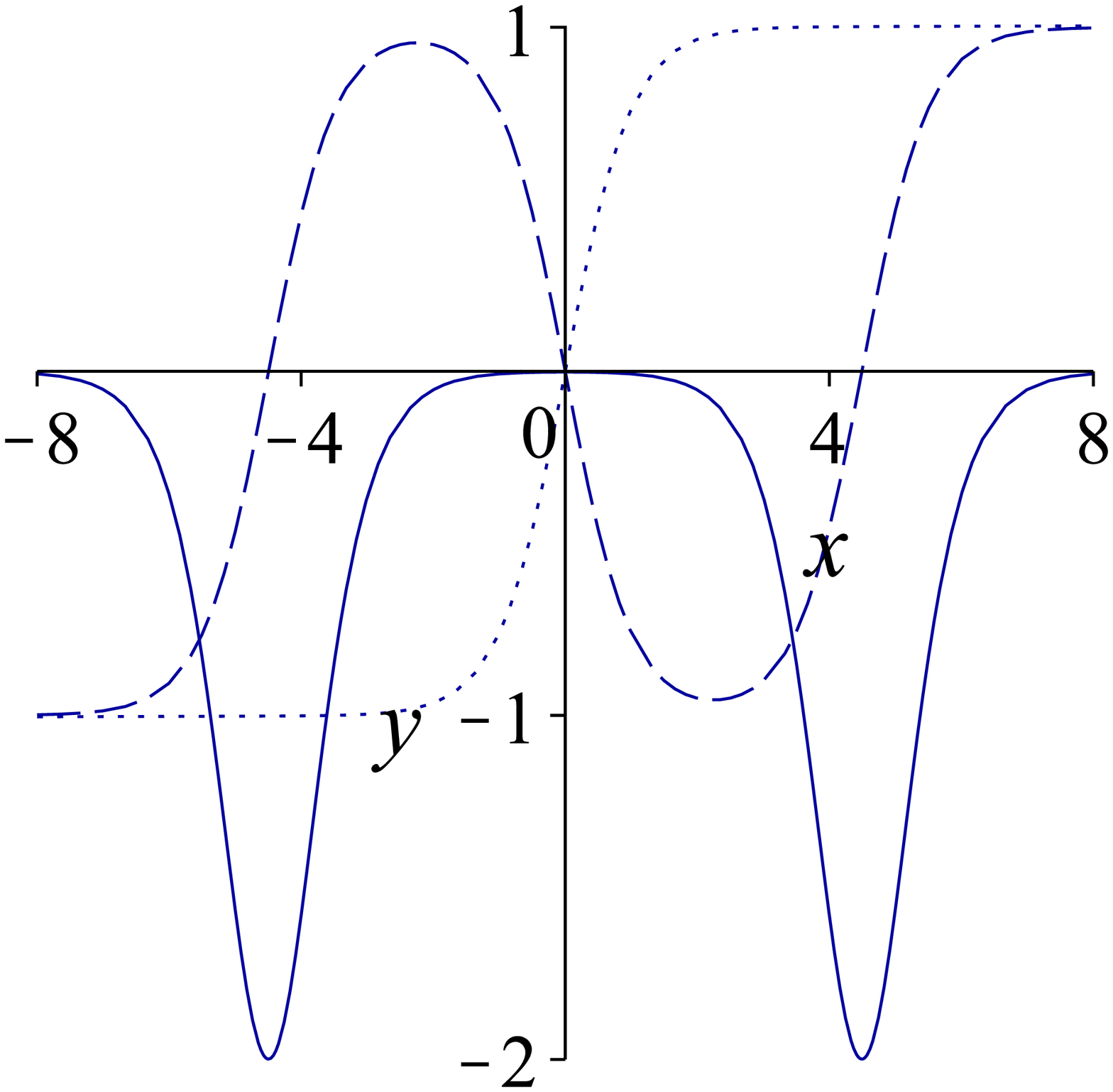}}\put(-20,20){\Large \bf a}
\subfigure{\label{fig:AsymmetricSuperpotentialPotentialPhi}
\includegraphics[width=0.23\textwidth]
{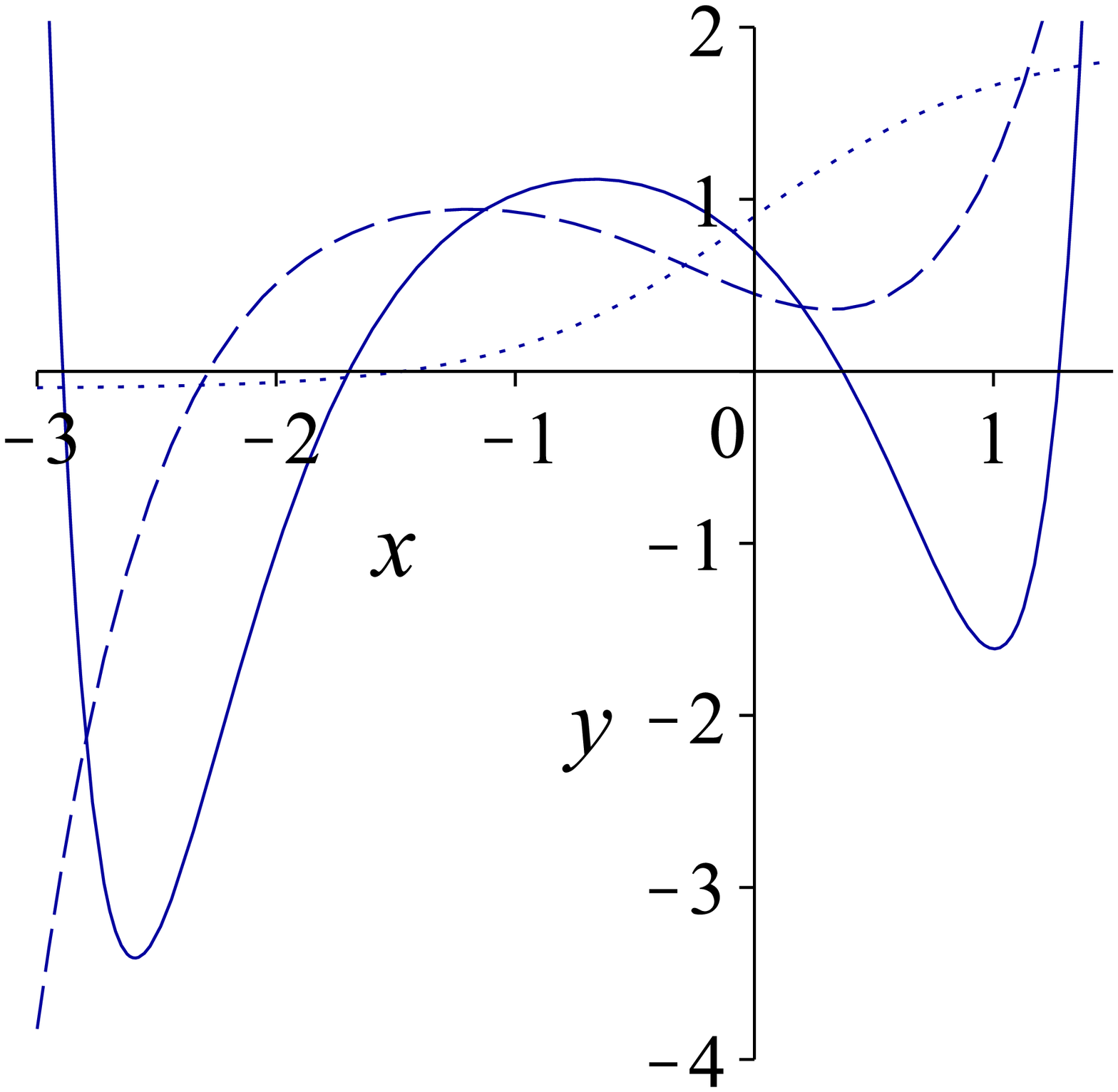}}\put(-20,20){\Large \bf b}
\caption{The \dblwel potential $V$ (solid line), the function $\chi$
 (dashed line) and the \multfun $\phi$ (dotted line).
 For the symmetric case, ({\bf a}, $\phi =
\sinh(ax)/ \cosh(bx)$, for the asymmetric case, ({\bf b}, $\phi =
\alpha + \tanh(\beta x)$. The symmetric case with parameters $a =
\sqrt{-E_0}$, $b = \sqrt{-E_1}$, $E_0 = -1$ and $E_1 = -0.999$,
features odd $\chi$ and $\phi$ functions, whereas for the asymmetric
case (here with $\alpha = 0.9$, $\beta = 1$, $E_0 = 0$ and $E_1 =
1$) these symmetries are lost.}
\label{fig:SuperpotentialPotentialPhiFunctions}
\end{figure}

The Schr\"odinger equation (for the two lowest states, $n = 0$ and
$n = 1$) is
\begin{equation}
\label{eq:SrhodingerEquation}
\frac{\hbar^2}{2m}{\psi_n}^{\prime\prime}(x) + \left[{E_n} - V(x)
\right]
{\psi_n}(x) = 0\; .
\end{equation}
Throughout this paper we choose $\hbar^2/2m = 1$. Similarly to
super-symmetric quantum mechanics~\cite{Gendenshtein_83,Cooper_95},
where a superpotential is defined in order to compute the \grndst and
the corresponding potential, in~\cite{Caticha_95} a \multfun $\phi$
is defined, which relates the ground and \frstxit states by
\begin{equation}
\label{eq:FunctionPhi}
\psi_1 = \phi\psi_0\; .
\end{equation}
Substituting Eq.~(\ref{eq:FunctionPhi}) into
Eq.~(\ref{eq:SrhodingerEquation}) for $n = 1$ yields
\begin{equation}
\label{eq:SEnequals1}
\left(\phi\psi_0\right)^{\prime\prime} + \left[E_1-V
\right]\phi\psi_0 = 0\; .
\end{equation}
Eq.~(\ref{eq:SrhodingerEquation}) for $n = 0$ multiplied by $\phi$ is
\begin{equation}
\label{eq:SEnequals0}
\phi\psi_0^{\prime\prime} + \left[E_0 - V\right]\phi\psi_0 = 0\; .
\end{equation}
Upon subtraction of Eq.~(\ref{eq:SEnequals0}) from
Eq.~(\ref{eq:SEnequals1}), one obtains a function $\chi$
\begin{equation}
\label{eq:Superpotential}
\chi(x) = \frac{\phi^{\prime\prime} + \Delta E\phi}{2\phi^\prime} =
-\frac{\psi_0^\prime}{\psi_0} = -\frac{d\ln\psi_0}{dx}\; ,
\end{equation}
where $\Delta E = E_1 - E_0$ is the energy splitting between ground
and \frstxit states. Eq.~(\ref{eq:Superpotential}) resembles a
superpotential in the context of super-symmetric quantum
systems~\cite{Gendenshtein_83,Cooper_95}. The corresponding \grndst
$\psi_0$ therefore is
\begin{equation}
\label{eq:Groundstate}
\psi_0(x) = \tildeN\exp\left(-\int_0^x\chi(x^\prime)dx^\prime
\right)\; ,
\end{equation}
here $\tildeN$ is a normalization constant. Rearranging
Eq.~(\ref{eq:SEnequals0}) determines the \dblwel potential $V$, up
to an additive constant
\begin{equation}
\label{eq:DoubleWellPotential}
V(x) = \frac{\psi_0^{\prime\prime}}{\psi_0} + E_0 = \chi^2 -
\chi^\prime + E_0\; .
\end{equation}
This is illustrated in
Fig.~\ref{fig:SuperpotentialPotentialPhiFunctions}.

\subsection{Symmetric \DblWel Potential}
\begin{figure}[ht]
\centering
\includegraphics[width=0.23\textwidth]{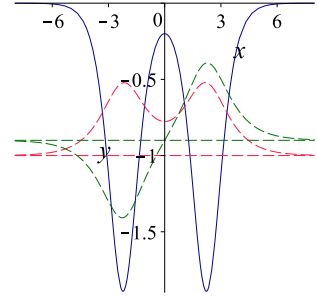}
\caption{(Color Online) The symmetric \dblwel potential $V$
 (solid blue line), the corresponding \grndst $\psi_0$
(dashed red line) and the \frstxit state $\psi_1$
(dashed green line), for $E_0 = -1$ and $E_1 = -0.9$.}
\label{fig:SymmetricDoubleWellPotential}
\end{figure}

For the case of a symmetric \dblwel potential, as displayed in
Figs.~\ref{fig:SuperpotentialPotentialPhiFunctions}~{\bf a}
and~\ref{fig:SymmetricDoubleWellPotential}, we use the \multfun
\begin{equation}
\label{eq:SymmetricCasePhiFunction}
\phi(x) = \frac{\sinh (ax)}{\cosh (bx)}\; ,
\end{equation}
where (with $E_0<E_1<0$)
\begin{equation}
\label{eq:SymmetricDoubleWellVariables} a = \sqrt{-E_0} \quad \mbox{
and } \quad b = \sqrt{-E_1}\; .
\end{equation}
The function $\chi$ then takes the form
\begin{multline}
\label{eq:SymmetricSuperpotential}
\chi(x) =\frac{\sinh (ax)\left(2b^2 - 2a^2\cosh^2 (bx)\right)}{b\sinh
(ax)\sinh (2bx)
- 2a\cosh (ax)\cosh^2 (bx)}\\
+ \frac{ab\cosh (ax)\sinh (2bx)}{b\sinh (ax)\sinh (2bx) - 2a\cosh
(ax)\cosh^2 (bx)}
\end{multline}
and the corresponding symmetric potential $V$ is
\begin{equation}
\label{eq:SymmetricPotential}
V(x) = \frac{2\left(b^2 - a^2\right)\left(a^2\cosh^2 (bx) +
b^2\sinh^2 (ax)\right)}{(a\cosh (ax)\cosh (bx) - b\sinh (ax)\sinh
(bx))^2}\; .
\end{equation}
The \grndst $\psi_0$ (note typographical error in
corresponding~{Eq.~{(19)}} of ref.~\cite{Caticha_95}), is
\begin{equation}
\label{eq:SymmetricGroundState} \psi_0(x) = \frac{\psi_0(0)(e^{2bx}
+ 1)(a - b)e^{ax}}{(e^{2(b + a)x} + 1)(a - b) + (a + b)(e^{2ax} +
e^{2bx})}
\end{equation}
and the \frstxit state $\psi_1$ is
\begin{equation}
\label{eq:SymmetricFirstExcited}
\psi_1(x) = \frac{\psi_1(0)(e^{2ax} - 1)(a - b)e^{bx}}{(e^{2(b +
a)x} + 1)(a - b) + (a + b)(e^{2ax} + e^{2bx})}\; .
\end{equation}
The two \wavfun{}s have odd and even parity, see
Fig.~\ref{fig:SymmetricDoubleWellPotential}.

\subsection{Asymmetric \DblWel Potential}
\begin{figure}[ht]
\centering
\includegraphics[width=0.23\textwidth]{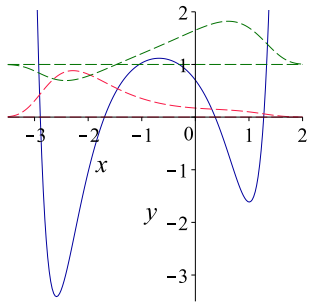}
\caption{(Color Online) The asymmetric \dblwel potential $V$ (solid
blue line), the corresponding \grndst $\psi_0$ (dashed red line) and
the \frstxit state~$\psi_1$ (dashed green line), for $\alpha = 0.9$,
$\beta = 1$, $E_0 = 0$ and $\Delta E = 1$.}
\label{fig:AsymmetricDoubleWellPotential}
\end{figure}

With the \multfun $\phi$
\begin{equation}
\label{eq:AsymmetricCasePhiFunction}
\phi(x) = \alpha + \tanh(\beta x)
\end{equation}
the function $\chi$ takes the form
\begin{equation}
\label{eq:AsymmetricSuperpotential}
\chi(x) = \frac{\Delta E}{4\beta}\left[\sinh(2\beta x) +
2\alpha\cosh^2(\beta x)\right] - \beta\tanh(\beta x)
\end{equation}
and the corresponding asymmetric potential $V$ is
\begin{multline}
\label{eq:AsymmetricPotential}
V(x) = \beta^2 - \Delta E\alpha\sinh (2\beta x)\\
+ \cosh^2 (\beta x)\left(\frac{\Delta E^2}{4\beta^2}\alpha\sinh
(2\beta x) - \frac{\Delta E^2}{4\beta^2} - 2\Delta E\right) \\
+ \frac{\Delta E^2}{4\beta^2}\left( \alpha^2 + 1\right)\cosh^4
(\beta x) + \frac{3}{2}\Delta E + E_0\; .
\end{multline}
The corresponding lowest two \wavfun{}s are
\begin{multline}
\label{eq:AsymmetricGroundState}
\psi_0(x) = \psi_0(0) \cosh (\beta x)\\
\times \exp\left[ -\frac{\Delta E}{4\beta^2}\left(\cosh^2 (\beta x)
+ \alpha\beta x + \frac{\alpha}{2}\sinh (2\beta x)\right)\right]
\end{multline}
and
\begin{multline}
\label{eq:AsymmetricFirstExcited}
\psi_1(x) = \psi_1(0) (\alpha\cosh (\beta x) + \sinh (\beta x))\\
\times \exp\left[ -\frac{\Delta E}{4\beta^2}\left(\cosh^2 (\beta x)
+ \alpha\beta x + \frac{\alpha}{2}\sinh (2\beta x)\right)\right]\; .
\end{multline}
Fig.~\ref{fig:AsymmetricDoubleWellPotential} illustrates that the
potential's asymmetry strongly modifies the shape of the lowest two
\wavfun{}s as compared to the symmetric case, with $\alpha = 0$,
in~Fig.~\ref{fig:SymmetricDoubleWellPotential}.

To investigate the tunneling dynamics we use the normalized
superposition state
\begin{multline}
\label{eq:Superposition} \Psi(x,t) =
\sin(\theta)\exp\left(\frac{-iE_0t}{\hbar}\right)\psi_0(x) \\
 + \cos(\theta)\exp\left(\frac{-iE_1t}{\hbar}\right)\psi_1(x)\; ,
\end{multline}
with the weighting angle $\theta \in [0,\ldots,\pi/2]$. The energy
splitting $\Delta E$, gives rise to the beat period (or reciprocal
barrier-tunneling rate~\cite{Merzbacher_02,Razavy_03}) $T =
2\pi\hbar/\Delta E$. In other words, $T$ is the time needed for a
quantum particle initially localized in, say, the left well, to
perform a full oscillation: left-right-left. The larger the energy
splitting the shorter the beat period.

\section{Wigner's Function}\label{sec:WignerFunction}

Eugene Wigner introduced his quasi-\probdist function $W(x,p;t)$ in
1932 for the study of quantum corrections to classical statistical
mechanics~\cite{Wigner_32}. It is a generating function for all
spatial auto-correlation functions of a given quantum mechanical
\wavfun $\psi$ and defined as~\cite{Razavy_03,Wigner_32,Belloni_04}
\begin{equation}
\label{eq:WignerDistributionFunction} W(x,p;t) =
\frac{1}{\pi\hbar}\int_{-\infty}^\infty{\Psi^*(x+y,t)\Psi(x-y,t)
e^{\frac{2ipy}{\hbar}}dy}\; ,
\end{equation}
where $x$ and $y$ are position variables and $p$ the momentum. The
Wigner quasi-\probdist function is a real-valued phase-space
distribution function, it can assume negative values which is why it
is referred to as a \emph{quasi}-\probdist.

Its marginals are the quantum-mechanical \probdist{}s of position
\begin{equation}
\label{eq:PositionProbabilityDistribution}
P(x,t) = \left|\Psi(x,t)\right|^2 = \int_{-\infty}^\infty
{W(x,p;t)\; dp}
\end{equation}
and momentum
\begin{equation}
\label{eq:MomentumProbabilityDistribution} \tilde P(p,t) =
\left|\Phi(p,t)\right|^2 = \int_{-\infty}^\infty {W(x,p;t)\; dx}\; .
\end{equation}
It is normalized $
\int_{-\infty}^\infty{\int_{-\infty}^\infty{W(x,p;t)\; dp}\; dx} = 1
\; $ and the overlap of the Wigner functions $W_\psi(x,p;t)$ and
$W_\chi(x,p;t)$ of two distinct quantum states, $\psi(x,t)$ and
$\chi(x,t)$, yields the magnitude of their \wavfun overlap
squared~\cite{Belloni_04}
\begin{equation}
\label{eq:WignerNonNegativeRelation}
\int_{-\infty}^\infty{\int_{-\infty}^\infty{W_\psi(x,p;t)
W_\chi(x,p;t)\; dp}\; dx} =
\frac{2}{\pi\hbar}\left|\left\langle\psi|\chi\right\rangle\right|^2.
\end{equation}

\begin{figure}[ht!]
\centering
\subfigure{\label{fig:SymmetricDoubleWellWignerDistribution1}
\includegraphics[width=0.23\textwidth]
{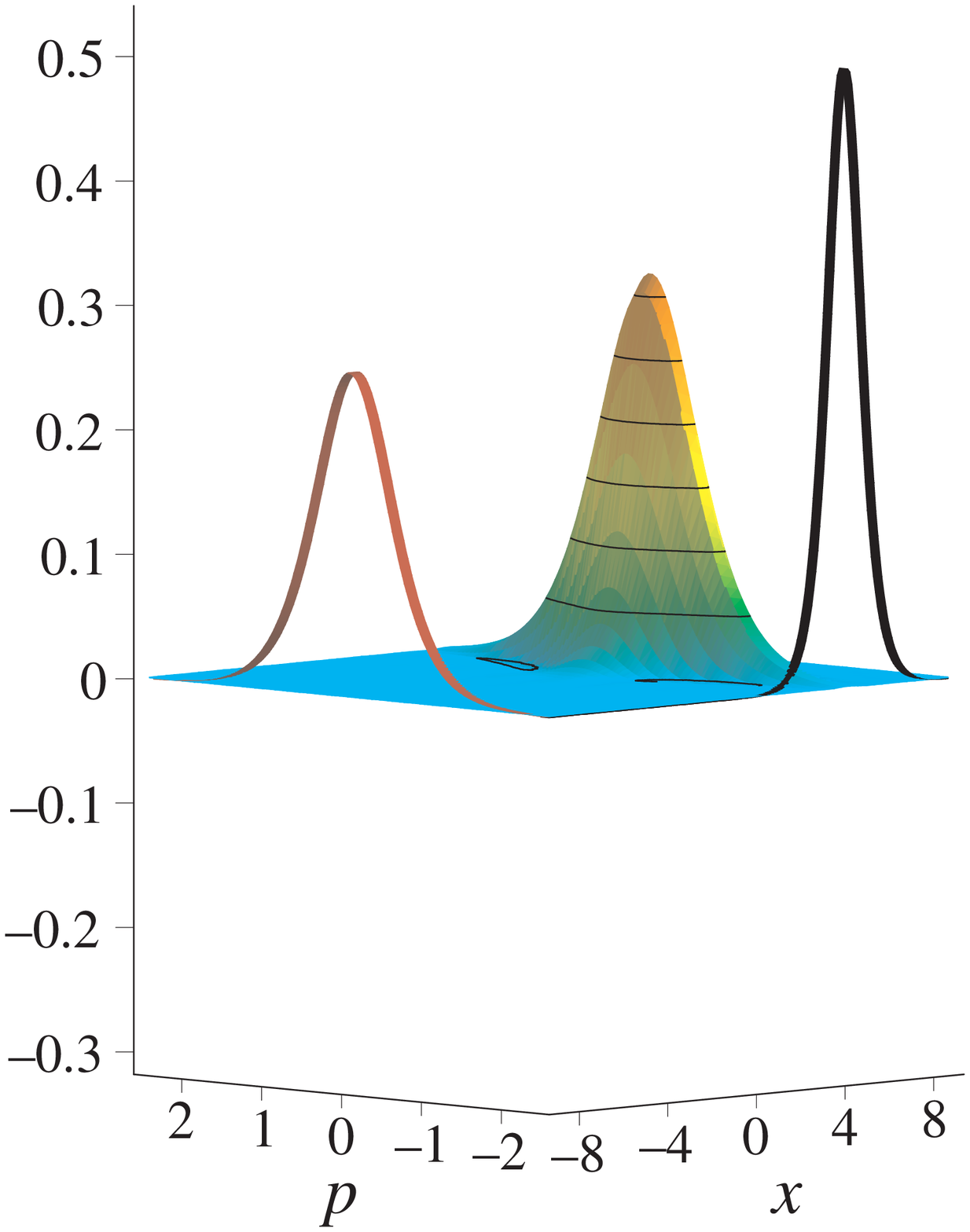}}
\put(-50,140){\Large \bf a}\put(-30,40){$\bf t=0$}
\subfigure{\label{fig:AsymmetricDoubleWellWignerDistribution1}
\includegraphics[width=0.23\textwidth]
{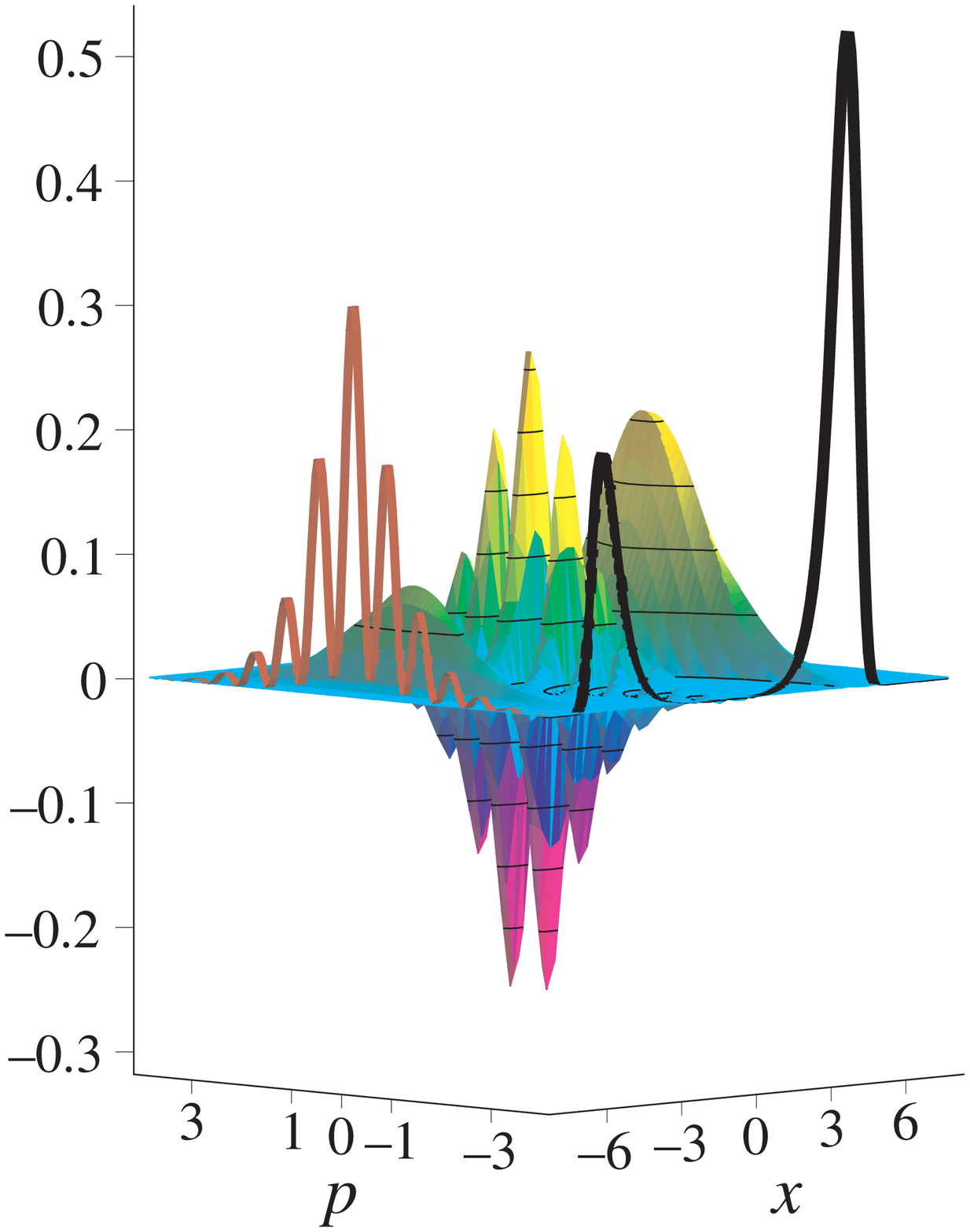}}
\put(-50,140){\Large \bf d}\put(-30,40){$\bf t=0$}\\
\subfigure{\label{fig:SymmetricDoubleWellWignerDistribution2}
\includegraphics[width=0.23\textwidth]
{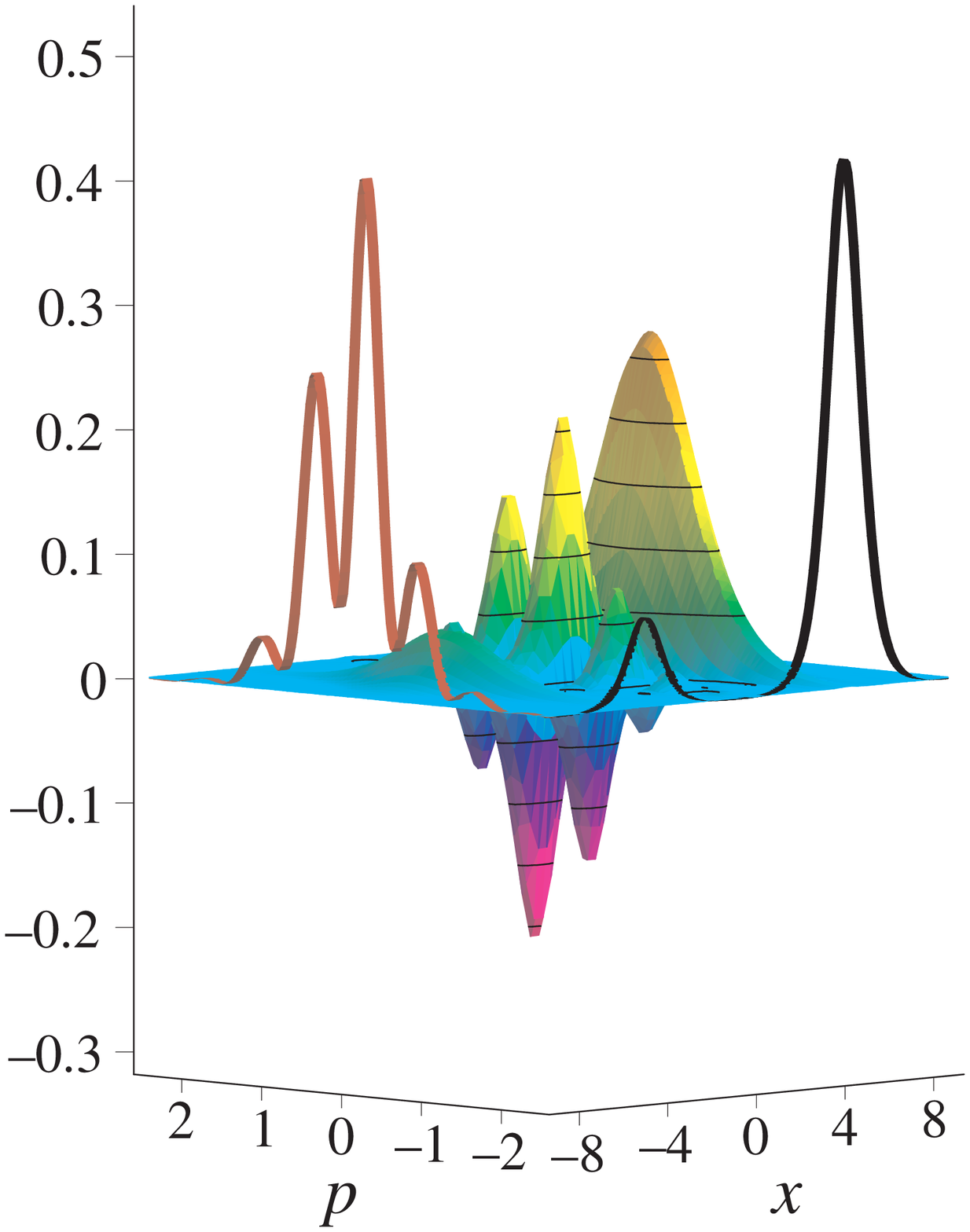}}
\put(-50,140){\Large \bf b}\put(-30,40){$\bf t=\frac{T}{8}$}
\subfigure{\label{fig:AsymmetricDoubleWellWignerDistribution2}
\includegraphics[width=0.23\textwidth]
{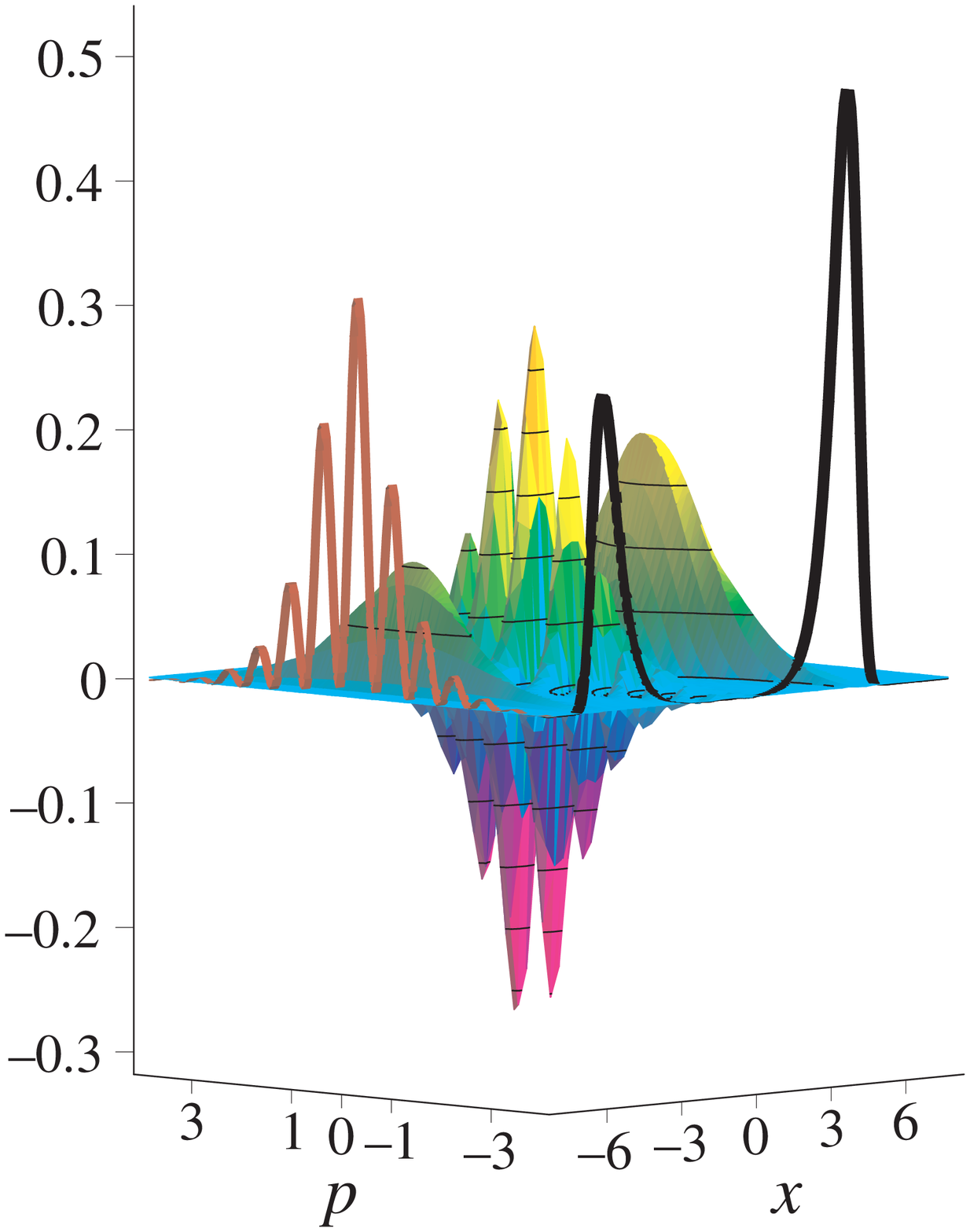}}
\put(-50,140){\Large \bf e}\put(-30,40){$\bf t=\frac{T}{8}$}\\
\subfigure{\label{fig:SymmetricDoubleWellWignerDistribution3}
\includegraphics[width=0.23\textwidth]
{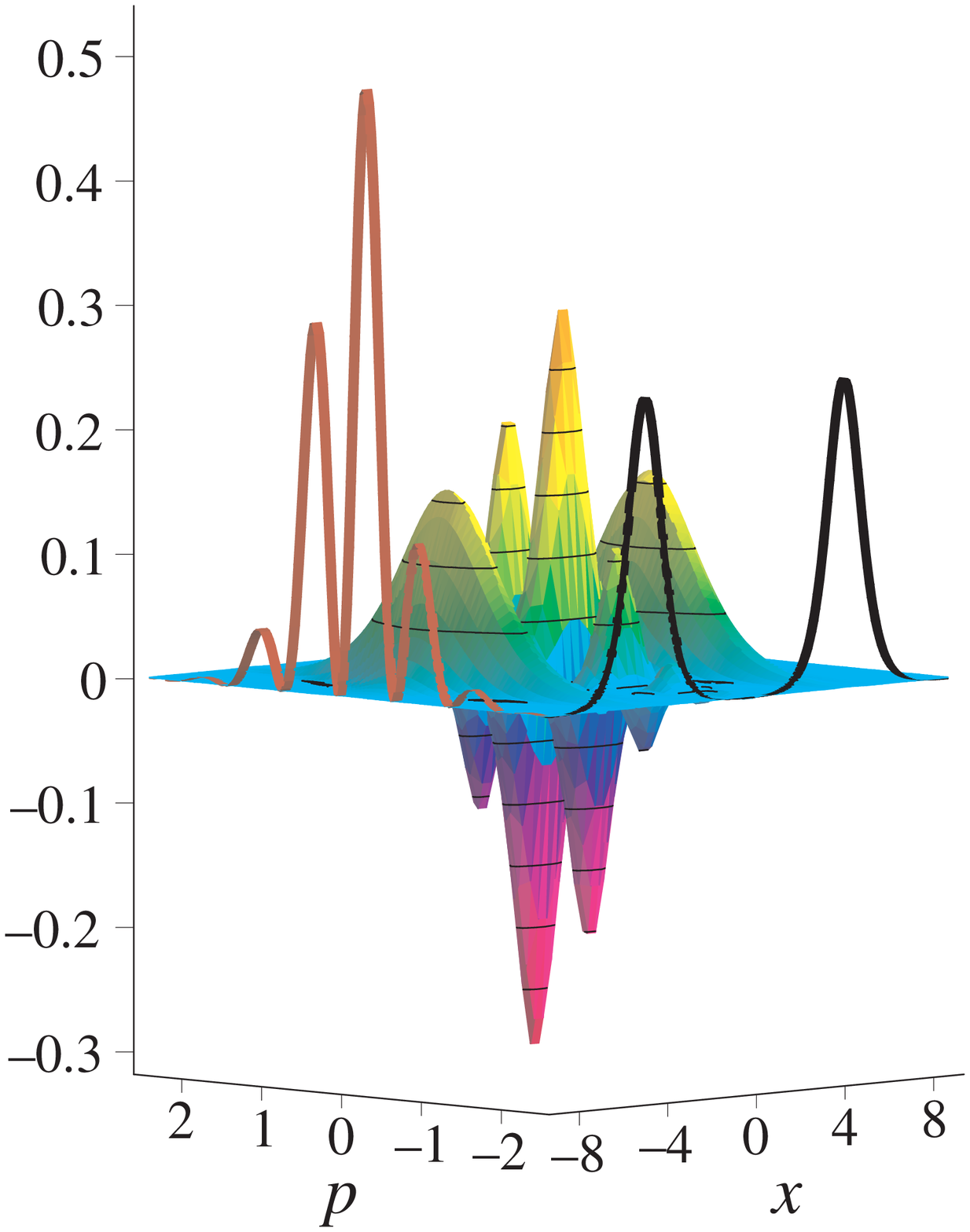}}
\put(-50,140){\Large \bf c}\put(-30,40){$\bf t=\frac{T}{4}$}
\subfigure{\label{fig:AsymmetricDoubleWellWignerDistribution3}
\includegraphics[width=0.23\textwidth]
{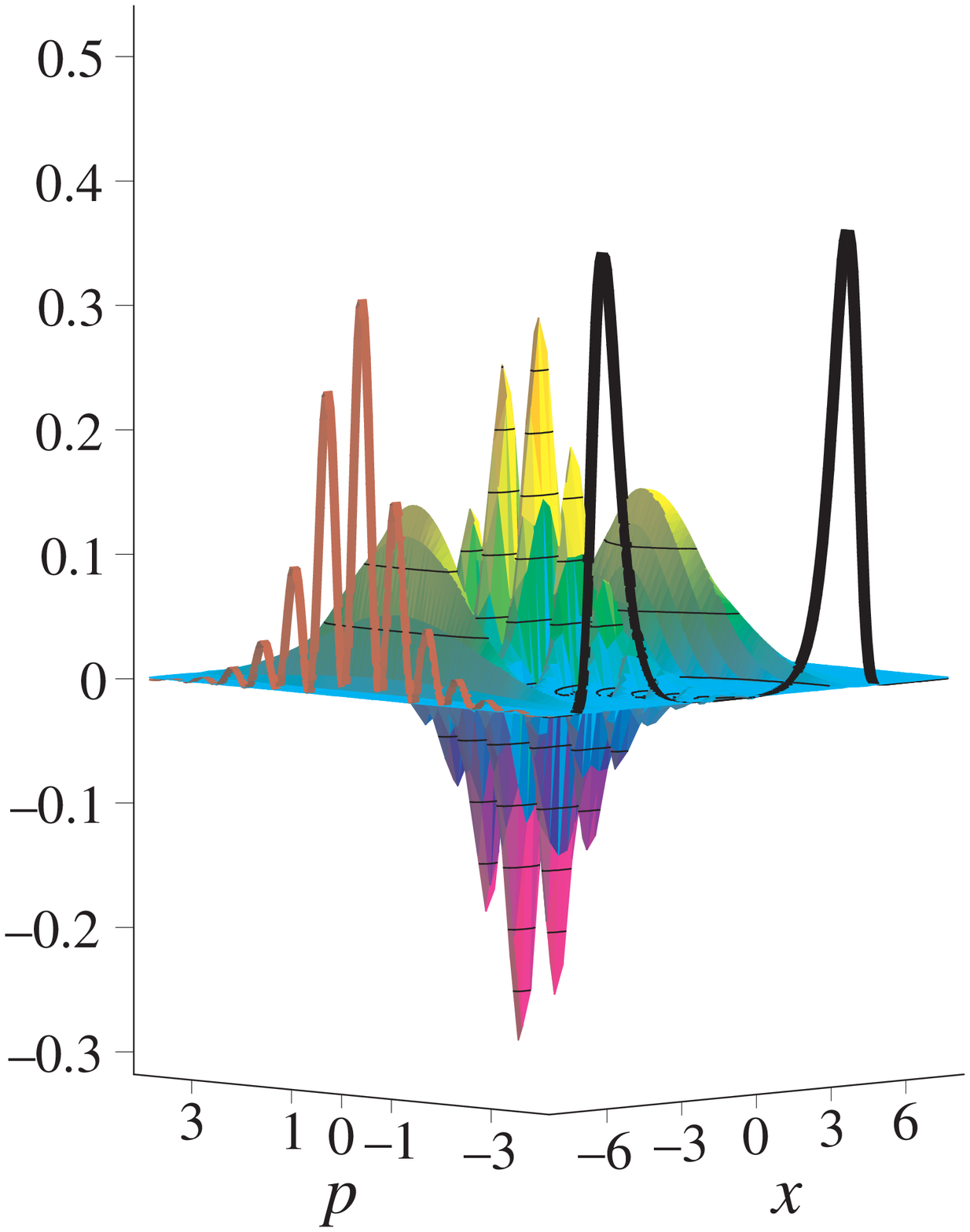}} \put(-50,140){\Large \bf
f}\put(-30,40){$\bf t=\frac{T}{4}$} \caption{(Color Online) Time
evolution of the Wigner quasi-\probdist, for the symmetric \dblwel
potential $V$ of~Eq.~(\ref{eq:SymmetricPotential}) for $E_0 = -1$
and $E_1 = -0.999$, shown in subfigures {\bf a}, {\bf b} and
{\bf c}, and for the asymmetric \dblwel potential $V$
of~Eq.~(\ref{eq:AsymmetricPotential}) for $\alpha = 0.9$, $\beta =
1$, $E_0 = 0$ and $E_1 = 0.001$, shown in subfigures {\bf d},
{\bf e} and {\bf f}, with weighting angle $\theta = \pi/4$. The
projections of the Wigner function are the \probdist{}s with respect
to position $P(x,t)$ (black) and momentum $\tilde P(p,t)/3$ (brown -
shrunk by ``$1/3$'' for clarity and visibility).}
\label{fig:DoubleWellWignerDistribution}
\end{figure}

Fig.~\ref{fig:DoubleWellWignerDistribution} shows plots of the time
evolution of the Wigner functions for symmetric and asymmetric \dblwel
potentials and the associated marginals $P(x,t)$ and $\tilde P(p,t)$;
all Wigner functions and marginals~$\tilde P(p,t)$ had to be
determined through numerical integrations.

\subsection{Negative values of Wigner function}\label{sec:QuantumCoherence}

As can be seen from the projections of the Wigner quasi-\probdist{}s
in Fig.~\ref{fig:DoubleWellWignerDistribution}, a particle that
exists in two places at the same time shows interference fringes in
its momentum \probdist $\tilde P(p,t)$. These are incompatible with
a single humped position distribution~$P(x,t)$ in conjunction with a
positive semi-definite phase space \probdist. A phase
space distribution simultaneously yielding such marginals has to
contain regions with negative values. These negative regions are
indicators of the non-classical character of the spatial coherences
of the \wavfun{}s~\cite{Schleich_01} and are frequently studied in
experiments~\cite{Grangier_11}. They arise, e.g., whenever the
\wavfun spreads out over both wells of the \dblwel potential.

To avoid confusion, we would like to emphasize that negative regions
of the Wigner function, as a generating function for spatial
auto-correlation functions, represent non-classical behavior of
spatial correlations but not of the non-classical behavior of
tunneling associated with ``negative kinetic energy". The interference
fringes of the Wigner function, appearing roughly in the area where
tunneling occurs, represent the non-classical spatial coherence of the
\wavfun{}s located in both wells simultaneously.


\begin{figure}[ht!]
\centering
\subfigure{\label
{fig:SymmetricDoubleWellWignerDistributionFringes1}
\includegraphics[width=0.23\textwidth]
{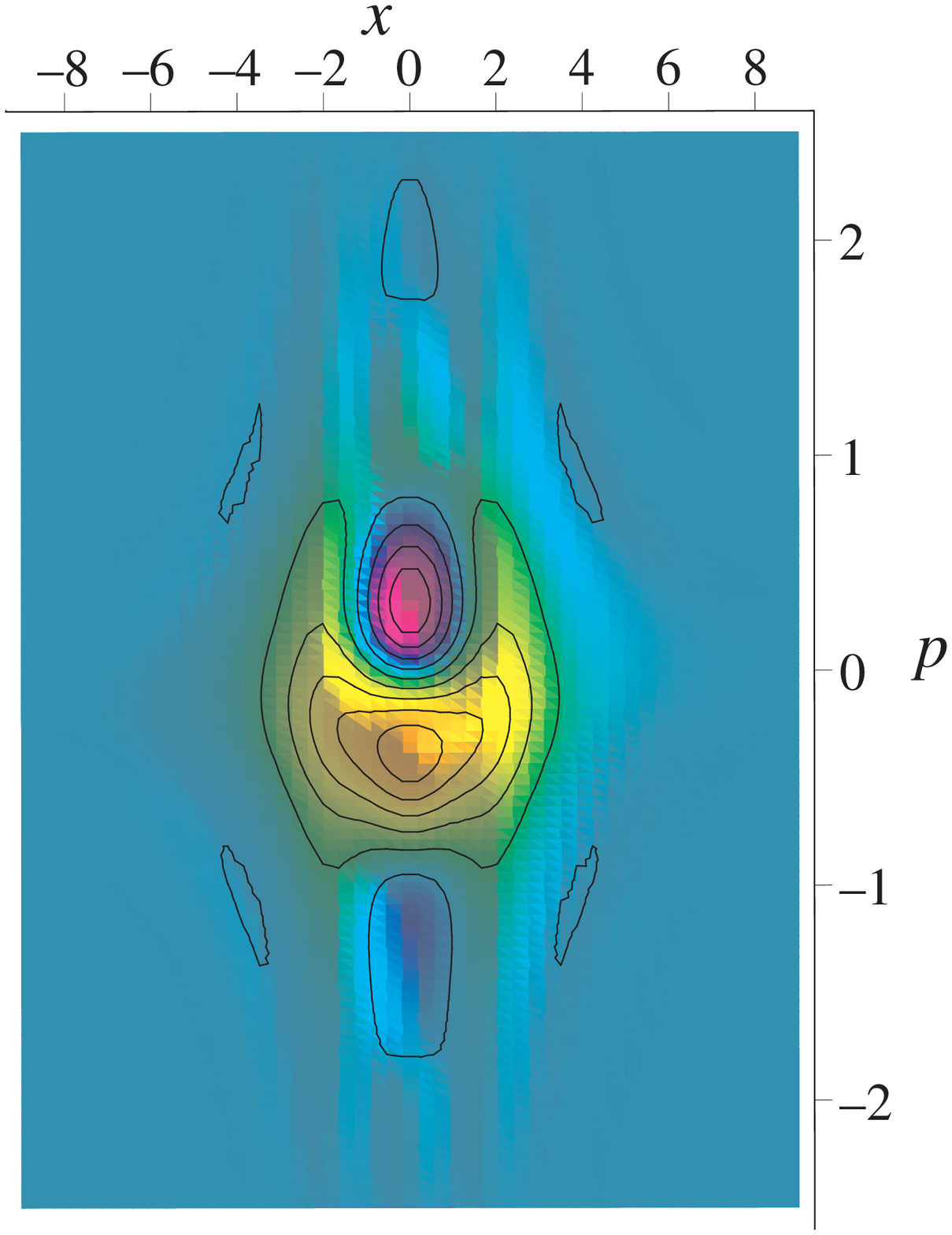}}
\put(-35,120){\Large \bf a}
\subfigure{\label
{fig:AsymmetricDoubleWellWignerDistributionFringes1}
\includegraphics[width=0.23\textwidth]
{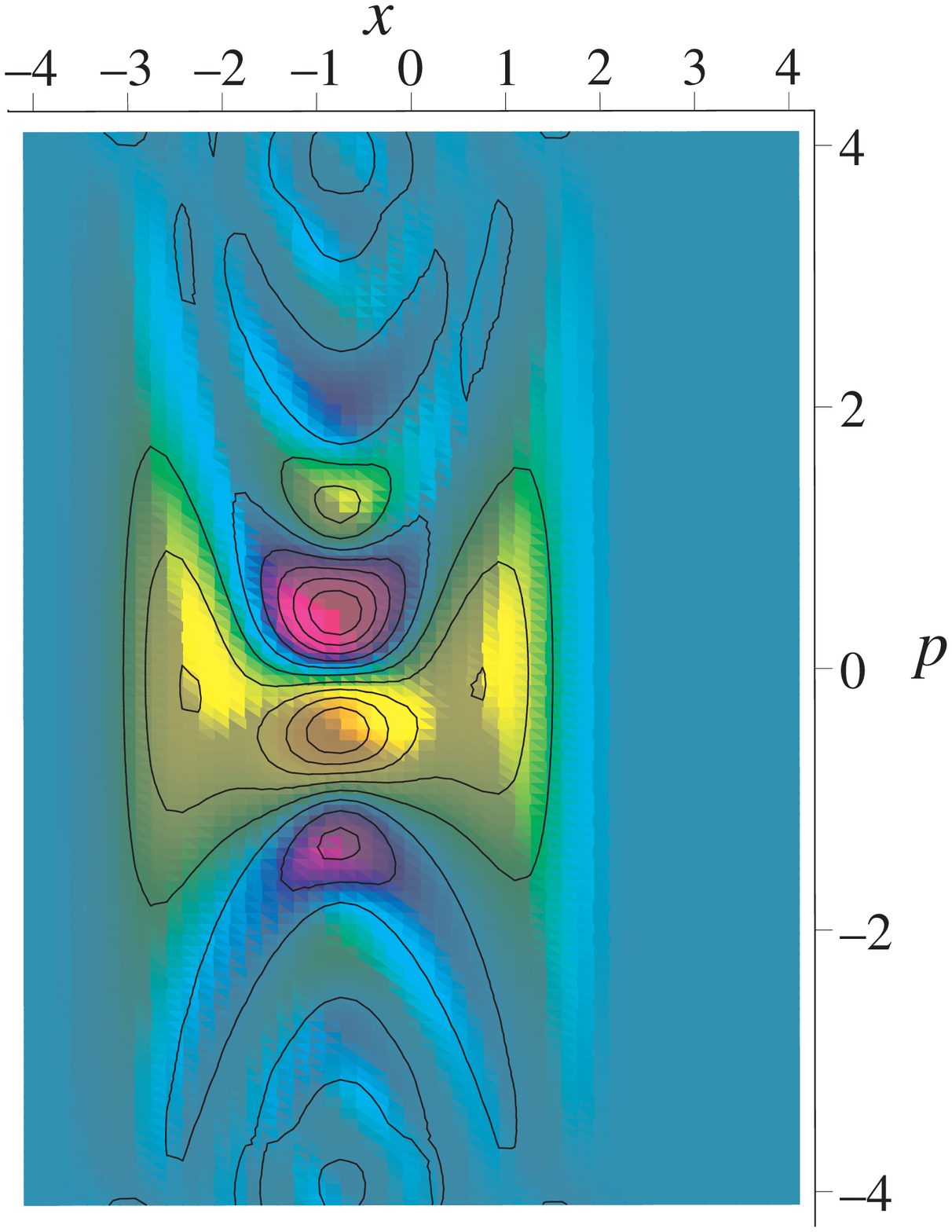}}
\put(-35,120){\Large \bf d}\\
\subfigure{\label
{fig:SymmetricDoubleWellWignerDistributionFringes2}
\includegraphics[width=0.23\textwidth]
{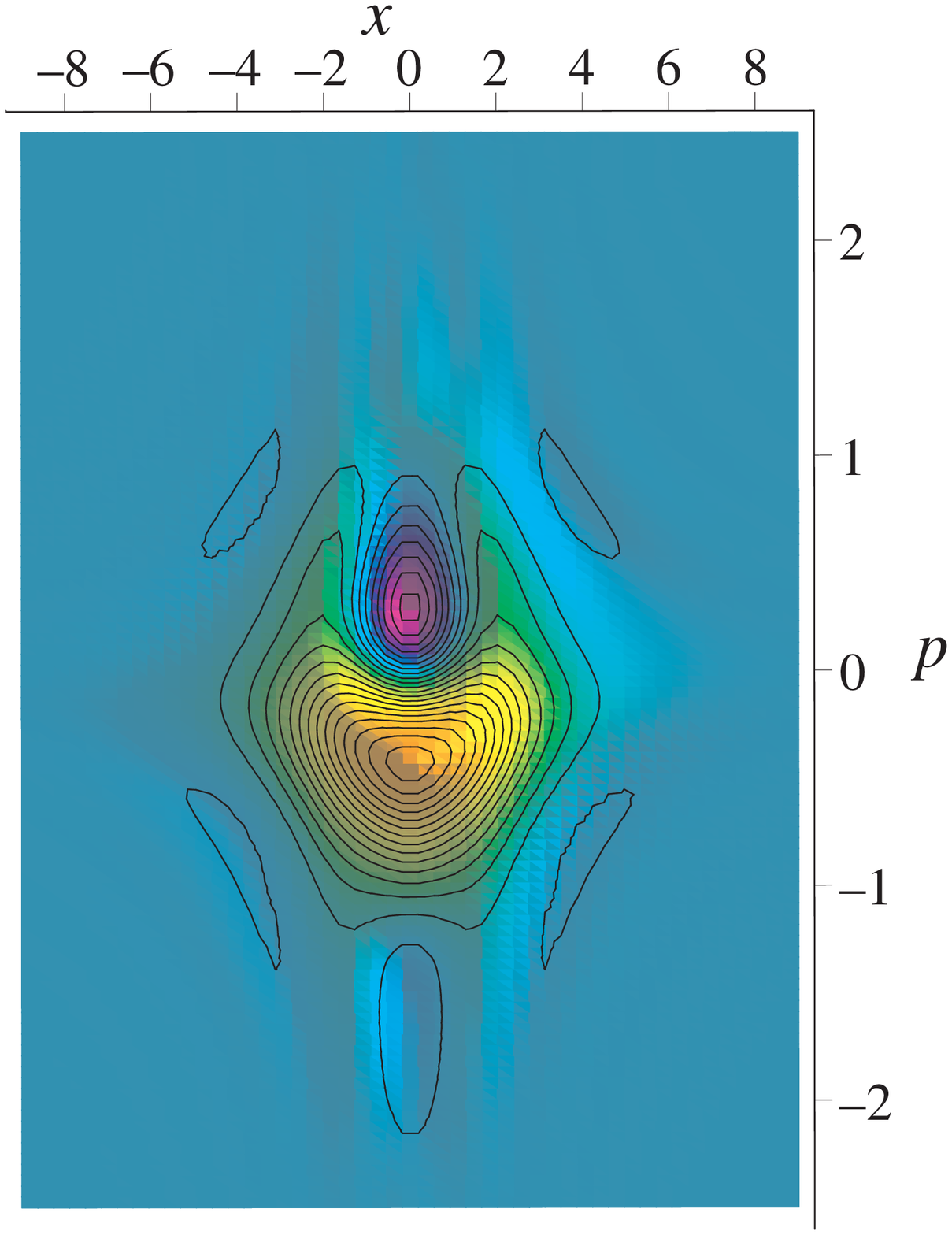}}
\put(-35,120){\Large \bf b}
\subfigure{\label
{fig:AsymmetricDoubleWellWignerDistributionFringes2}
\includegraphics[width=0.23\textwidth]
{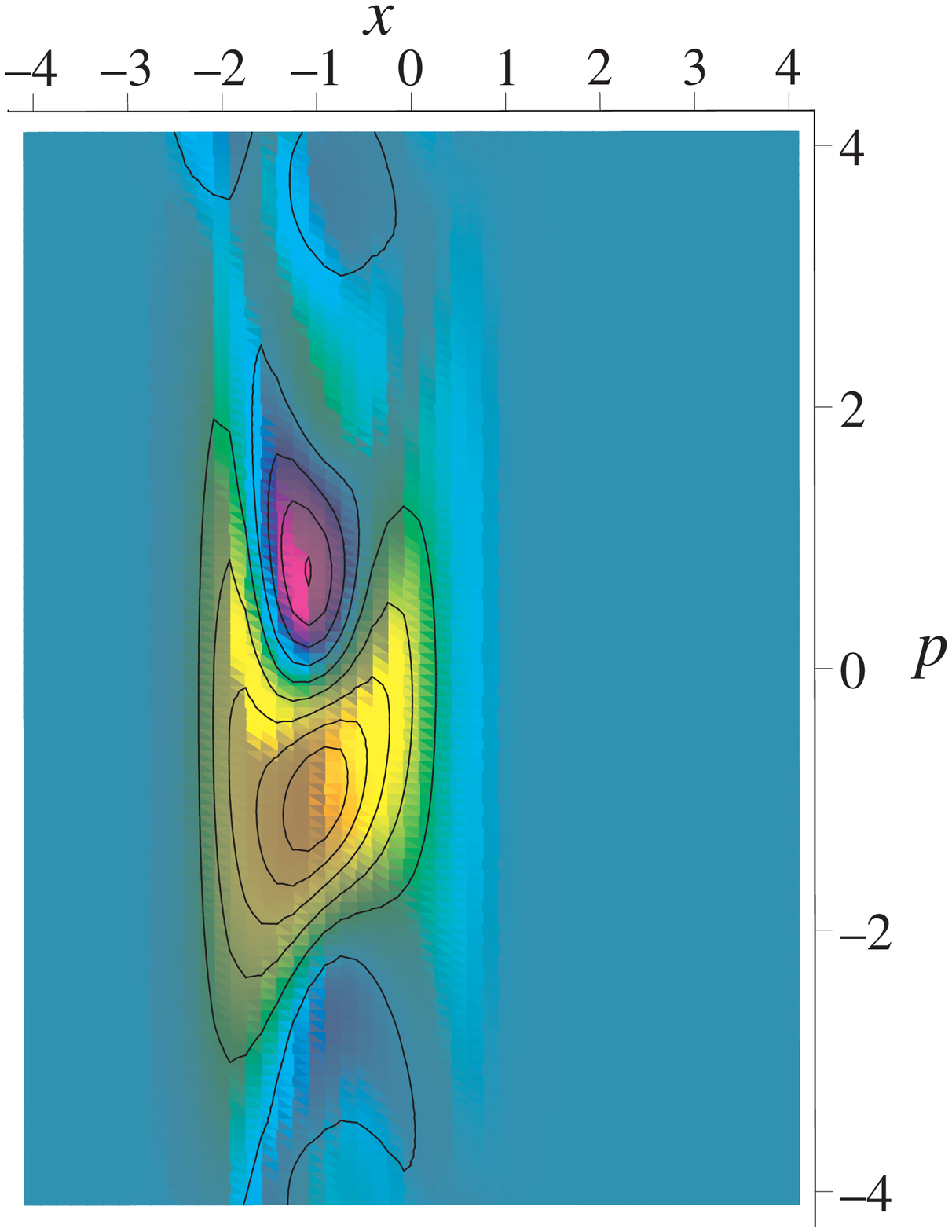}}
\put(-35,120){\Large \bf e}\\
\subfigure{\label
{fig:SymmetricDoubleWellWignerDistributionFringes3}
\includegraphics[width=0.23\textwidth]
{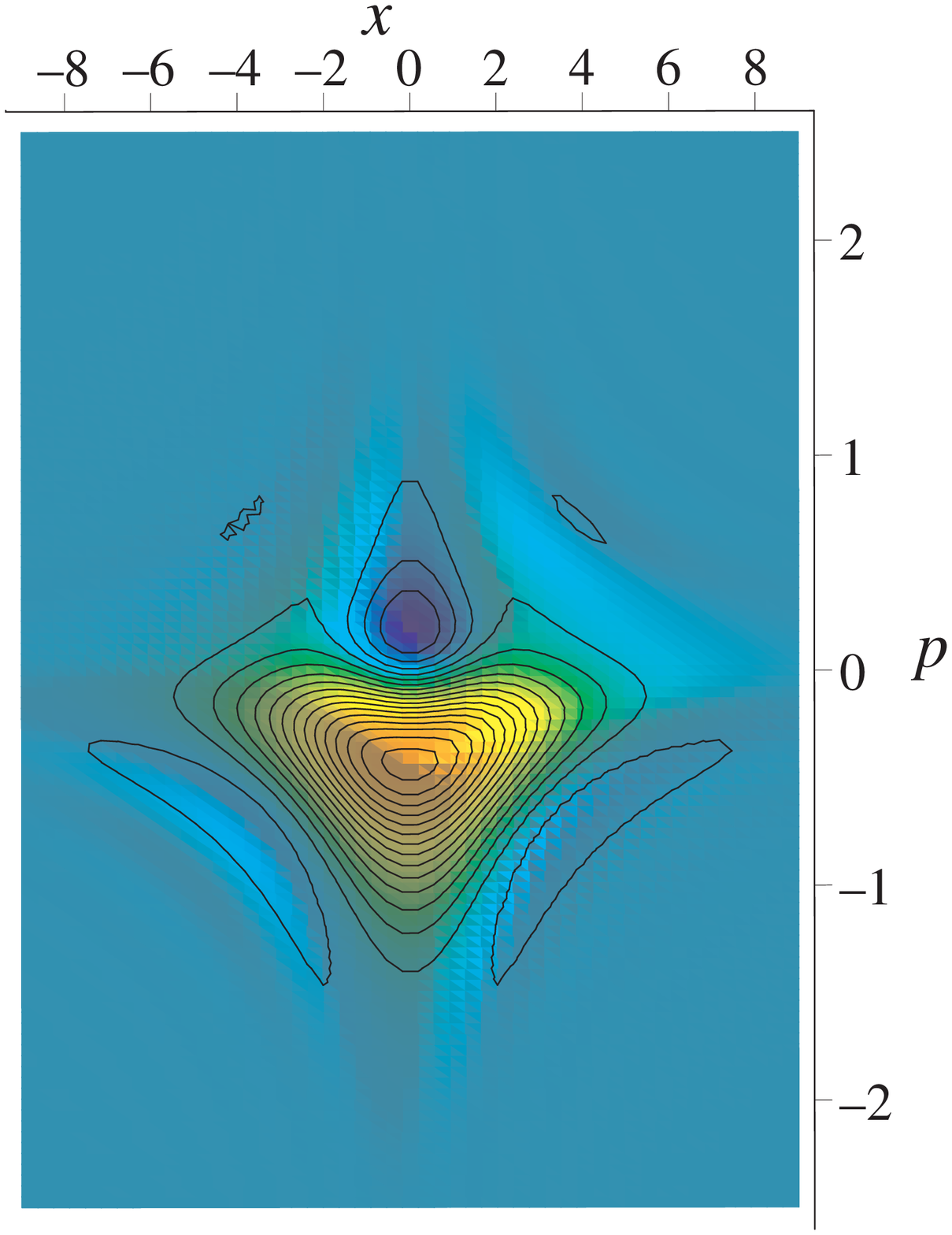}}
\put(-35,120){\Large \bf c}
\subfigure{\label
{fig:AsymmetricDoubleWellWignerDistributionFringes3}
\includegraphics[width=0.23\textwidth]
{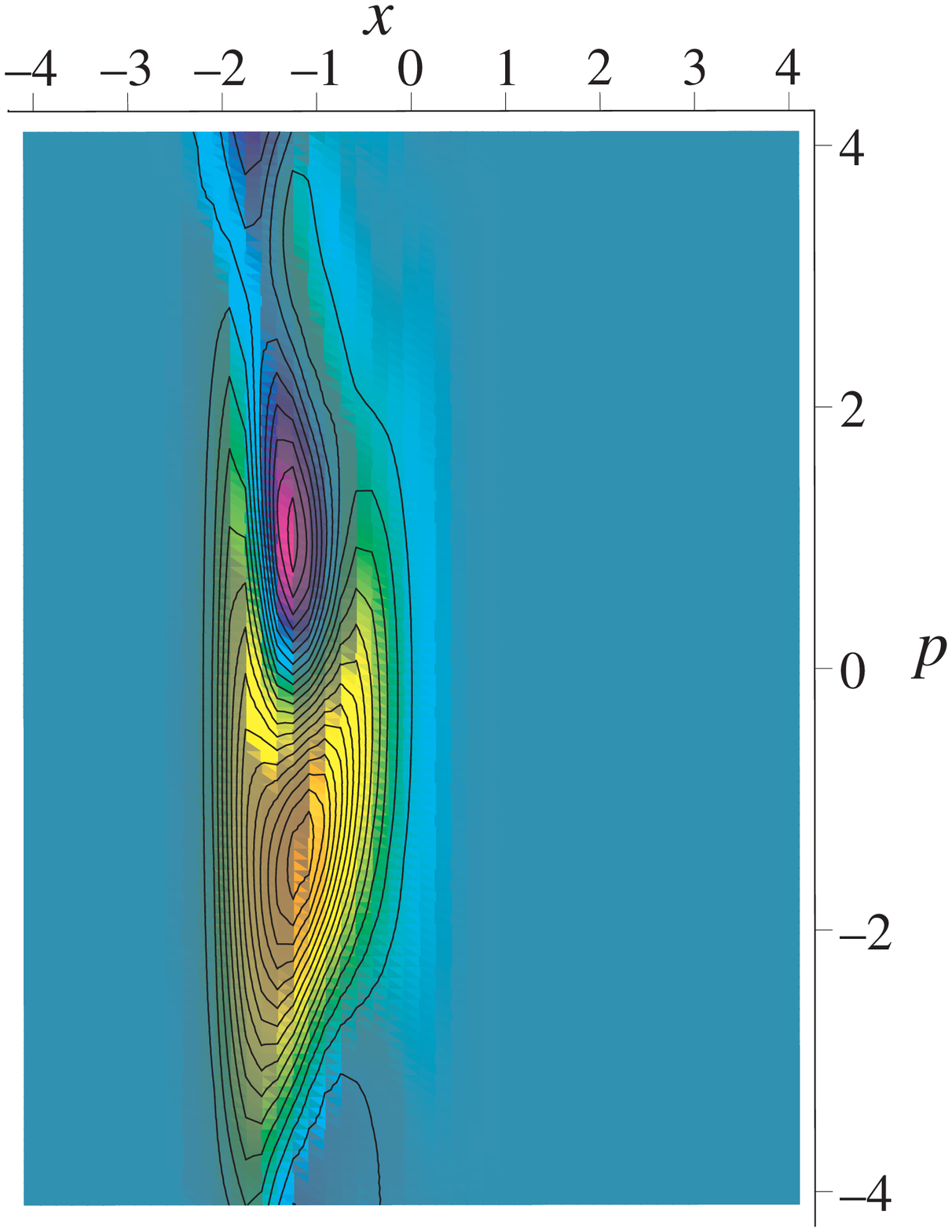}} \put(-35,120){\Large
\bf f} \caption{(Color Online - Color scheme identical to
Fig.~\ref{fig:DoubleWellWignerDistribution}'s) The Wigner
quasi-\probdist for a range of values of the energy splitting $\Delta
E$, for the symmetric \dblwel potential $V$
of~Eq.~(\ref{eq:SymmetricPotential}) for $E_0 = -1$ and $\Delta E =
0.25, 0.5\mbox{ and } 0.75$, shown in subfigures {\bf a}, {\bf b} and
{\bf c}, respectively, and for the asymmetric \dblwel potential $V$
of~Eq.~(\ref{eq:AsymmetricPotential}) for $\alpha = 0.9$, $\beta = 1$,
$E_0 = 0$ and for $\Delta E = 0.5, 4\mbox{ and } 8$, shown in
subfigures {\bf d}, {\bf e} and {\bf f}, respectively, with weighting
angle $\theta = \pi/4$ and $t=T/4$.}
\label{fig:DoubleWellWignerDistributionFringes}
\end{figure}

\begin{figure}[ht!]
\centering
\subfigure{\label{fig:SymmetricDoubleWellPotentialRangeE1}
\includegraphics[width=0.23\textwidth]
{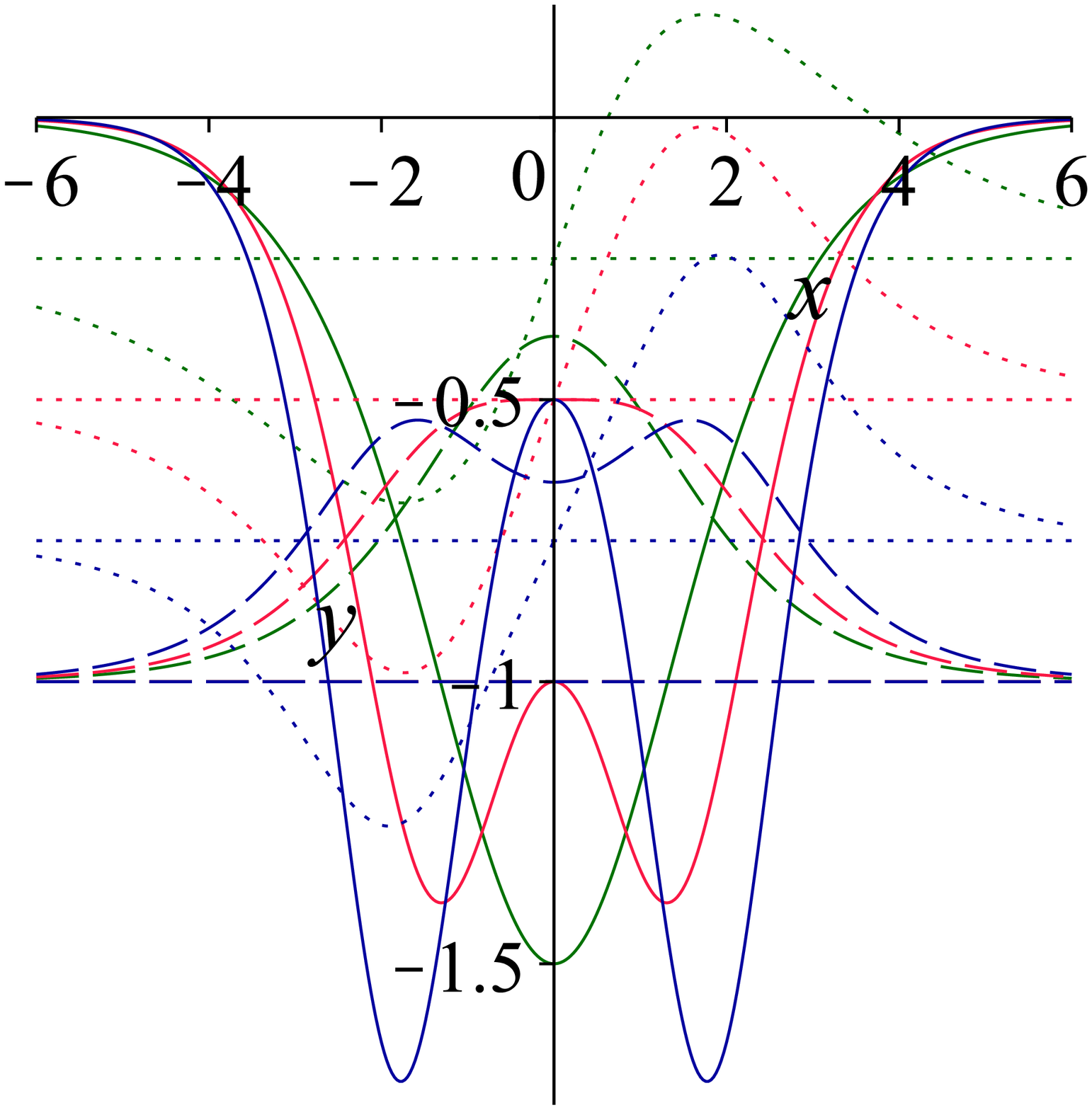}}\put(-20,20){\Large \bf a}
\subfigure{\label{fig:AsymmetricDoubleWellPotentialRangeDeltaE}
\includegraphics[width=0.23\textwidth]
{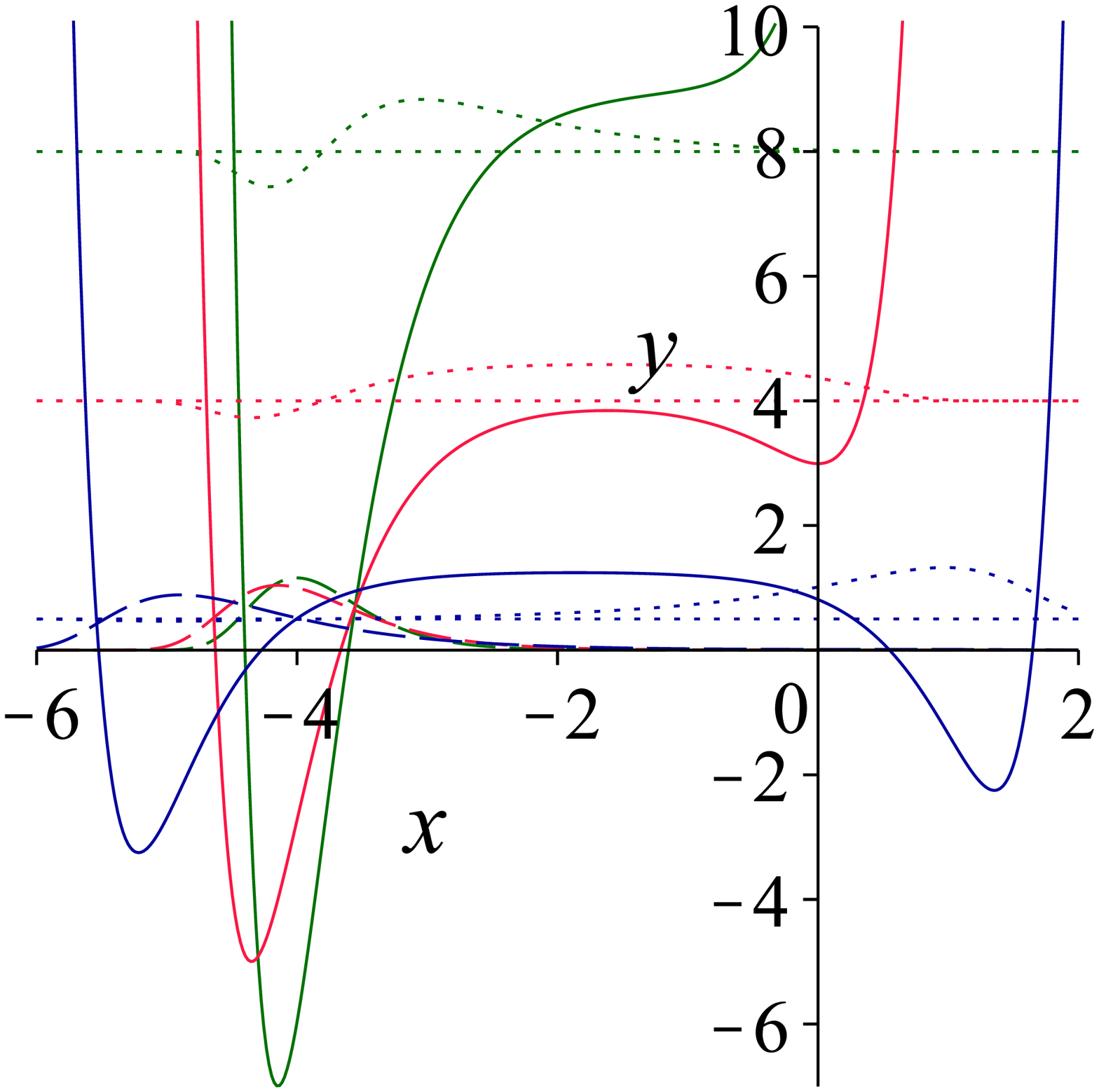}}\put(-20,20){\Large \bf
b} \caption{(Color Online) The symmetric \dblwel potential $V$ of
Eq.~(\ref{eq:SymmetricPotential}), {\bf a}, for $E_0=-1$ and
$\Delta E = 0.25, 0.5\mbox{ and } 0.75$, color blue, red and green,
respectively, and the asymmetric \dblwel potential $V$ of
Eq.~(\ref{eq:AsymmetricPotential}), {\bf b}, for $\alpha = 0.9$,
$\beta = 1$, $E_0 = 0$ and $\Delta E = 0.5, 4\mbox{ and } 8$, color
blue, red and green, respectively, and their corresponding ground
and \frstxit states, $\psi_0$ and $\psi_1$.}
\label{fig:DoubleWellPotentialRangeDeltaE}
\end{figure}

Fig.~\ref{fig:DoubleWellWignerDistributionFringes} illustrates the
changes Wigner's functions undergo when changing the potential wells
from separate wells to merged single troughs (displayed in
Fig.~\ref{fig:DoubleWellPotentialRangeDeltaE}): Since the coupling
between the wells increases, so does $\Delta E$. Concurrently the
peaks of the spatial wave functions move together which increases
the fringe spacing in the associated momentum representation,
visible as a reduction of the spatial frequency of Wigner's
functions' interference patterns.

\section{Conclusion}

We have considered the effects of tunneling in smooth partially
exactly solvable \dblwel potentials~\cite{Caticha_95}. We illustrated
the behavior of the associated \wavfun{}s and Wigner's
quasi-probability phase-space distributions. Wigner's functions can
assume negative values representing non-classical spatial coherences
of the \wavfun{}s, these were shown to arise in the case of tunneling
and analyzed and interpreted.

\pagebreak

%


\end{document}